\documentclass{elsart}

\usepackage{graphicx}
\usepackage{amsmath}
\usepackage{natbib}
\usepackage{booktabs}

\begin{document}

\begin{frontmatter}

\title{A New Time-Symmetric Block Time-Step Algorithm for N-Body Simulations}

\author{Murat Kaplan\thanksref{1}}, 
\author{Hasan Sayg\i n\thanksref{2}},
\author{Piet Hut\thanksref{3}},
\author{Jun Makino\thanksref{4}}

\thanks[1]{Istanbul Technical University, Informatics Institute, Computational 
Science and Engineering Program,  34469 Maslak, Istanbul, Turkey}
\thanks[2]{Istanbul Technical University, Institute of Energy, 34469 Maslak, 
Istanbul, Turkey}
\thanks[3]{Institute for Advanced Study, Princeton, NJ 08540, USA}
\thanks[4]{Department of Astronomy, University of Tokyo, 7-3-1 Hongo,
  Bunkyo-ku, Tokyo 113-0033, Japan}

\begin{abstract}
Time-symmetric integration schemes share with symplectic schemes the
property that their energy errors show a much better behavior than is
the case for generic integration schemes.  Allowing adaptive time
steps typically leads to a loss of symplecticity.  In contrast, time
symmetry can be easily maintained, at least for a continuous choice of
time step size.  In large-scale N-body simulations, however, one often
uses block time steps, where all time steps are forced to take on
values as powers of two.  This greatly facilitates parallelization,
and hence code efficiency.  Straightforward implementation of
time-symmetry, translated to block time steps, faces significant
hurdles.  For example, iteration can lead to oscillatory behavior,
and even when such behavior is suppressed, energy errors show a linear
drift in time.  We present an approach that circumvents these problems.
\end{abstract}

\begin{keyword}
N-body simulations\sep Celestial mechanics\sep Stellar dynamics

\PACS 95.10.Ce\sep 98.10.+z
\end{keyword}
\end{frontmatter}
\maketitle

\section{Introduction}

For long-term N-body simulations, it is essential that the drift in
the values of conserved quantities is kept to a minimum.  The total
energy is often used as an indicator of such a drift.  During the
last fifteen years, two approaches have been put forward to improve
numerical conservation of energy and other theoretically conserved
quantities: symplectic integration schemes, where the simulated system
is guaranteed to follow a slightly perturbed Hamiltonian system, and
time-symmetric integration schemes, where the simulated system follows
the same trajectory in phase space, when run backwards or forwards

In both cases, for symplectic as well as for time-symmetric schemes,
the introduction of adaptive time steps tends to destroy the desired
properties.  Symplectic schemes are perturbed to different Hamiltonians
at different choices of time step length, and therefore lose their
global symplecticity.  Time-symmetric schemes typically determine
their time step length at the beginning of a step, which implies that
running a step backwards gives a slightly different length for that step.
See \citet{Leimkuhler-2005} for a recent review of various attempts to
remedy this situation.

In practice, many large-scale simulations in stellar dynamics use a
block time-step approach, where the only allowed values for the time
step length are powers of two \citep{Aarseth-2003}.  The name derives
from the fact that, with this recipe, many particles will share the
same step size, which implies that their orbit integration can be
performed in parallel.  An added benefit, in the case of individual
time steps, is the fact that block time steps allow one to predict
the positions of all particles only once per block time step, rather
than separately for each particle that needs to be moved forward.
Since parallelization is rapidly becoming essential for any major
simulation, we explore in this paper the possibility to extend time
symmetry to the use of block time steps.

In Section 2, we analyze some of the problems that occur when applying
existing methods to the case of block time steps, and we offer a novel
solution, with a truly time symmetric choice of time step, with the
restriction that we only allow changes of a factor two in the
direction of increasing and decreasing the time step.
In Section 3, we present numerical tests of our new scheme, which show
the superiority of our approach over various alternatives.
Section 4 sums up.

\section{Block Time Steps}

\subsection{Implicit Iterative Time Symmetrization}
\label{subsec:implicit}

It is surprisingly easy to introduce a time-symmetric version for any
adaptive self-starting integration scheme.  Let $\xi = (r,v)$ be the
$2N$-dimensional phase space vector for a system with $N$ degrees of
freedom, and let $f(\xi_i,\delta t_i)$ be the operator that maps the
phase space vector of the system at time $t_i$ to a new phase space
vector at time $t_{i+1} = t_i + \delta t_i$.  Any choice of
self-starting integration scheme, together with a recipe to determine
the next time step $\delta t_i$, at time $t_i$ and phase space value
$\xi_i(t_i)$, defines the precise form of $f(\xi_i,\delta t_i)$.

The recipe for making any such scheme time-symmetric was given by
\citet{Hut-1995}, as:


\begin{equation}
\left\{ \begin{array}{lcl}
\xi_{i+1} &=& f(\xi_i,\delta t_i),\\
\phantom{1}&\phantom{1}&\phantom{1} \\
\delta t_i &=&
{\displaystyle \frac{h(\xi_i)+h(\xi_{i+1})}{2}} \label{time_symm}
\end{array} \right.
\end{equation}

where $h(\xi)$ can be any time step criterion.  Note that this recipe
leads to an implicit integration scheme, which can be solved most
easily through iteration.  In practice, one or two iterations suffice
to get excellent accuracy, but at the cost of doubling or tripling the
number of force calculations that need to be performed.  Extensions
of this implicit symmetrization idea have been presented by
\citet{Funato-1996} and \citet{Hut-1997}.

Since we will need to inspect the idea of iteration below in more detail,
let us write out the process here explicitly.  We start with the given
state $\xi_i$ and the implicit equation for $\xi_{i+1}$ of the form

\begin{equation}
\xi_{i+1} = f(\xi_i,\delta t_i(\xi_i, \xi_{i+1}))
\end{equation}

The first guess for $\xi_{i+1}$ is

\begin{equation}
\xi_{i+1}^{(0)} = f(\xi_i,\delta t_i(\xi_i, \xi_i))
\end{equation}

and we can consider this as our zeroth-order iteration.  With this
guess in hand, we can now start to iterate, finding

\begin{equation}
\xi_{i+1}^{(1)} = f(\xi_i,\delta t_i(\xi_i, \xi_{i+1}^{(0)}))
\end{equation}

as our first-order iteration.  This will already be much closer to the
final value, as long as the time steps are small enough and the
function $\delta t_i$ does not fluctuate too rapidly.  In general, the
$k^{th}$ iteration will yield a value for $\xi_{i+1}$ of

\begin{equation}
\xi_{i+1}^{(k)} = f(\xi_i,\delta t_i(\xi_i, \xi_{i+1}^{(k-1)}))
\end{equation}

We will now consider the application of these techniques to block time
steps.  For the purpose of illustrating the use of block time steps,
it will suffice to use the leapfrog scheme (also known as the
Verlet-St\"ormer-Delambre scheme, according to the authors who
rediscovered this scheme at roughly century-long intervals), which we
present here in a self-starting, but still time-symmetric form:

\begin{eqnarray}
r_{i+1}&=& r_{i} + v_{i}\delta t + a_i(\delta t)^2/2 \nonumber\\
v_{i+1} &=& v_i + (a_i + a_{i+1}) \delta t/2	 \label{leapfrog}
\end{eqnarray}

\subsection{Flip-Flop Problem}
\label{subsec:flipflop}

To start with, we apply the recipe of \citet{Hut-1995} to block time steps.
Let us define a block time step at level $n$ as having a length:

\begin{equation}
\Delta t_n = \frac{\Delta t_1}{2^{n-1}}. \label{block_time}
\end{equation}

where $\Delta t_1$ is the maximum time step length.
Starting with the continuum choice of

\begin{equation}
\delta t_{c,i} = \frac{h(\xi_i)+h(\xi_{i+1})}{2}
\end{equation}

we now force each time step to take on the block value
$\delta t_i = \Delta t_n$ for the smallest $n$ value
that obeys the condition
$\Delta t_n \le \delta t_{c,i}$.
In more formal terms,
$\delta t_i = \delta t_i(\delta t_{c,i}) = \Delta t_n$
for the unique $n$ value for which

\begin{equation}
n = \min_{k \ge 1}
\left\{
 k \ \Big| \ \frac{1}{2^{k-1}} \le \frac{h(\xi_i)+h(\xi_{i+1})}{2}
\right\}                                           \label{blockcondition}
\end{equation}

The problem with this approach is that we are no longer guaranteed to
find convergence for our iteration process, as can be seen from the
following example.  Let $h(\xi_i) = 0.502$ and let the time derivative
of $h(\xi_i(t))$ along the orbit be $(d/dt)h(\xi_i(t)) = -0.01$.  We
then get the following results for our attempt at iteration.

\begin{eqnarray}
\xi_{i+1}^{(0)} &=& f(\xi_i,\delta t_i(\xi_i, \xi_i))
= f(\xi_i,\delta t_i(h(\xi_i)))                        \nonumber\\
&=& f(\xi_i,\delta t_i(0.502))     
= f(\xi_i, 0.5)
\end{eqnarray}

\begin{eqnarray}
\xi_{i+1}^{(1)} &=& f(\xi_i,\delta t_i(\xi_i, \xi_{i+1}^{0}))
= f(\xi_i,\delta t_i([h(\xi_i) + h(\xi_{i+1}^{0})]/2))   \nonumber\\
&=& f(\xi_i,\delta t_i([0.502 + (0.502+0.5*(-0.01))]/2))         \nonumber\\
&=& f(\xi_i,\delta t_i([0.502 + 0.497]/2))                       \nonumber\\
&=& f(\xi_i,\delta t_i(0.4995))       
= f(\xi_i, 0.25)
\end{eqnarray}

\begin{eqnarray}
\xi_{i+1}^{(2)} &=& f(\xi_i,\delta t_i(\xi_i, \xi_{i+1}^{1}))
= f(\xi_i,\delta t_i([h(\xi_i) + h(\xi_{i+1}^{1})]/2))        \nonumber\\
&=& f(\xi_i,\delta t_i([0.502 + (0.502+0.25*(-0.01))]/2))         \nonumber\\
&=& f(\xi_i,\delta t_i([0.502 + 0.4995]/2))                       \nonumber\\
&=& f(\xi_i,\delta t_i(0.50075))        
= f(\xi_i, 0.5)
\end{eqnarray}

And from here on, $\xi_{i+1}^{(k)} = f(\xi_i, 0.25)$ for every odd value of $k$
and $\xi_{i+1}^{(k)} = f(\xi_i, 0.5)$ for every even value of $k$: the process
of iteration will never converge.

Under realistic conditions, for slowly varying $h$ functions and small
time steps, this flip-flop behavior will not occur often, but it will
occur sometimes, for a non-negligible fraction of the time.  We can
see this already from the above example: for a linear decrease in the
$h$ function of $(d/dt)h(\xi_i(t)) = -0.01$, we will get flip-flopping
not only for $h(\xi_i) = 0.502$ but for any value in the finite range
$ 0.50125 < h(\xi_i) < 0.5025$.

Since iteration converges correctly over the rest of the interval
$ 0.5 < h(\xi_i) < 1$, we conclude that in this particular case
flip-flopping occurs about one quarter of one percent of the time,
over this interval.  This is far too frequent to be negligible in a
realistic situation.

Clearly, a straightforward extension of the implicit iterative time
symmetrization approach does not work for block time steps, because
iteration does not converge.  We have to add some feature, in some
way.  Our first attempt at a solution is to take the smallest of the
two values in a flip-flop situation.

\subsection{Flip-Flop Resolution}
\label{subsec:noflipflop}

The most straightforward solution of the flip-flop dilemma is like
cutting the Gordian knot: we just take the lowest value of the two
alternate states.  The drawback of this solution is that in general
we need at least two iterations for each time step, to make sure that we
have spotted, and then correctly treated, all flip-flop situation.
In general, it is only at the third iteration that it becomes obvious
that a flip-flop is occurring.  To see this, consider the previous
example with a starting value of $h(\xi_i) = 0.501$.
In that case we will get $\xi_{i+1}^{(0)} = f(\xi_i, 0.5)$ and
$\xi_{i+1}^{(1)} = f(\xi_i, 0.25)$, just as when we started with
$h(\xi_i) = 0.502$.
The difference shows up only at the second iteration, where we now
find $\xi_{i+1}^{(2)} = f(\xi_i, 0.25)$, a value that will hold for
all higher iterations as well.

The original iterative approach to time symmetry in practice already
gives good results when we use only one iteration.  This implies a
penalty, in terms of force calculations per time steps, of a factor
two compared to non-time-symmetric explicit integration.  Now the
use of flip-flop resolution will force us to always take at least two
iterations per step, raising the penalty to become at least a factor
of three.

However, there is a more serious problem: there is still no guarantee
that taking the lowest value in a flip-flop situation leads to a
time-symmetric recipe.  In fact, what is even more important, we have
not yet checked whether our symmetric block time-step scheme is really
time symmetric, in the absence of flip-flop complications.

In order to investigate these questions, let us return to the example
we used above, but instead of a linear time derivative, let us now use
a quadratic time derivative for the $h$ function that gives the
estimate for the time step size.  Rather than writing a formal
definition, let us just state the values, while shifting the time
scale so that $t=0$ coincides with the particle position being
$\xi_i$:

\begin{eqnarray}
h(0.00) &=& 0.502 \nonumber\\
h(0.25) &=& 0.499 \nonumber\\
h(0.50) &=& 0.499
\end{eqnarray}

When we start at time $t=0$, and we integrate forwards. we find:

\begin{eqnarray}
\xi_{i+1}^{(0)} &=& f(\xi_i,\delta t_i(\xi_i, \xi_i))
= f(\xi_i,\delta t_i(h(\xi_i)))                        \nonumber\\
&=& f(\xi_i,\delta t_i(0.502))     
= f(\xi_i, 0.5)
\end{eqnarray}

\begin{eqnarray}
\xi_{i+1}^{(1)} &=& f(\xi_i,\delta t_i(\xi_i, \xi_{i+1}^{0}))
= f(\xi_i,\delta t_i([h(\xi_i) + h(\xi_{i+1}^{0})]/2))   \nonumber\\
&=& f(\xi_i,\delta t_i([0.502 + 0.499]/2))                       \nonumber\\
&=& f(\xi_i,\delta t_i(0.5005))       
= f(\xi_i, 0.5)
\end{eqnarray}

and so on: all further $k$th iterations will result in
$\xi_{i+k}^{(1)} = f(\xi_i, 0.5)$.  There is no flip-flop situation,
when moving forward in time.

However, when we now turn the clock backward, after taking this step
of half a time unit, we start with the value $h(0.50) = 0.499$, which
leads to a first step back of $\delta t = 0.25$.  The end point of the
first step back is $t = 0.25$ with $h(0.25) = 0.499$.  Therefore, also
here there is no flip-flop situation: all iterations, while going
backward, result in a time step size of $\delta t = 0.25$.

We have thus constructed a counter example, where forward integration
would proceed with time step $\delta t = 0.5$ and subsequent backward
integration would proceed with time step $\delta t = 0.25$.  Clearly,
our scheme is not yet time symmetric, even in the absence of a
flip-flop case.

\subsection{A First Attempt at a Solution}
\label{subsec:firsttry}

Let us rethink the whole procedure.  The basic problem has been that
the very first step in any of our algorithms proposed so far has not
been time symmetric.  The very first step moves forward, and leads to
a newly evolved system at the end of the first step.  Only {\it after}
making such a trial integration, do we look back, and try to restore
symmetry.  However, as we have seen, the danger is large that this
trial integration is not exhaustive: it may already go too far, or not
far enough, and thereby it may simply overlook a type of move that the
same algorithm would make if we would start out in the time-reversed
direction.

Formulating the problem in this way, immediately suggests a solution.
At any point in time, let us first try to make the largest step that
is allowed.  If that step turns out to be too large for our algorithm,
we try a step that is half that size.  if that step is too large still,
we again half the size, and so on, until we find a step size that
agrees with our algorithm, {\it when evaluated in both time directions.}
A similar treatment has been described by Quin et al (1997).

This type of approach is clearly more symmetric than what we have
attempted so far.  Instead of using information of the physical system
at the starting point of the next integration step, we only use a
mathematical criterion to find the largest time step size allowed at
that point, {\it and we then apply the physical criteria symmetrically
in both directions.}

Let us give an example.  If the largest time step size is chosen to be
unity, then at time $t = 0$ we start by considering this time step.
We try, in this order $\delta t = 1$, $\delta t = 0.5$, $\delta t = 0.25$,
and so on, until we find a time step for which integration starting in
the forward direction, and integration starting in the backward direction,
both result in the new time step being acceptable.  Let us say that
this is the case for $\delta t = 0.125$.

After taking this step, we are at time $t = 0.125$.  The largest time
step allowed at that point, forward or backward, is $\delta t = 0.125$.
Any larger time step would result in non-alignment of the block time
steps: in the backward direction it would jump over $t = 0$.  So at
this point we start by considering once more $\delta t = 0.125$.  If
that time step is too large, we try half that time step, halving it
successively until we find a satisfactory time step size.

Imagine that the second time step size is also $\delta t = 0.125$.  In
that case, we land at $t = 0.25$.  From there on, the maximum allowed
time step size is $\delta t = 0.25$, so the first try should be that
size.

In principle, this approach seems to be really time symmetric.
However, there is a huge problem with this type of scheme, as we have
just formulated it.  Imagine the system to crawl along with time steps
of, say $\delta t = 1/1024$, and reaching time $t = 1$.  Our new recipe
then suggests to start by trying $\delta t = 1$, a 1024-fold increase
in time step!  Whatever subtle physical effect it was that forced us
to take such small time steps, is completely ignored by the mathematical
recipe that forces us to look at such a ridiculously large time step.

For example, in the case of stellar dynamics, a double star may force
the stars that orbit each other to take time steps that are necessarily
far shorter than the orbital period.  Starting out with a trial step size
that is far larger than an orbital period may or may not give spuriously
safe-looking results.  Clearly, we have to exclude such enormous jumps in
time step.

\subsection{A Second Attempt at a Solution}
\label{subsec:secondtry}

The simplest solution to taming sudden unphysical increases in time
steps is to allow at most an increase of a factor two, in either the
forward or the backward direction.  This then implies that we can only
allow decreases of a factor two, and not more than two, in either
direction.  The reason is that a decrease of a factor four in one
direction in time would automatically translate into an increase of a
factor four in the other direction.

Note that we have to be careful with our time step criterion.  If we
allow time steps that are too large, we may encounter situations where
our time step criterion would suggest us to shrink time steps by a
factor of four, from one step to the other.  Since our algorithm does
not allow this, we can at most shrink by a factor of two, which may
imply an unacceptably large step.  However, if our time step criterion
is sufficiently strict, allowing only reasonably small time steps too
start with, it will be able to resolve the gradients in the criterion
in such as way as to handle all changes gracefully through halving and
doubling.

When we apply this restriction to the scheme outlined in the previous
subsection, we arrive at the following compact algorithm.

First a matter of notation.  Any block time step, of size $\delta t = 1/2^k$,
connects two points in time, only one of which can be written at
$t = Z/2^{(k-1)}$, with $Z$ an integer.  Let us call that time value an
{\it even time}, from the point of view of the given time step size, and
let us call that other time value an {\it odd time}.  To give an example, if
$\delta t = 0.125$, than $t = 0, 0.25, 0.5, 0.75, 1$ are all even times,
while $t = 0.125, 0.375, 0.625, 0.875$ are all odd times.

Here is our algorithm:

\begin{itemize}
\item[]
When we start in a given direction in time, at a given point
in time, we should determine the time step size of the last step made
by the system.  In that way, we can determine whether the current time
is even or odd, with respect to that last time step.

\medskip

\item[]
If the current time is odd, our one and only choice is: to continue
with the same size time step, or to halve the time step.  First, we
try to continue with the same time step.  If, upon iteration, that
time step qualifies according to the time-symmetry criterion used
before, Eq. \ref{blockcondition}, we continue to use the same time
step size as was used in the previous time step.  If not, we use half
of the previous time step.

\medskip

\item[]
If the current time step is even, we have a choice between three
options for the new time step size: doubling the previous time step
size, keeping it the same, or halving it.  We first try the largest
value, given by doubling.  If Eq.\ref{blockcondition} shows us that
this larger time step is not too large, we accept it, otherwise we
consider keeping the time step size the same.  If Eq.\ref{blockcondition}
shows us that keeping the time step size the same is okay, we accept
that choice, otherwise we just halve the time step, in which case no
further testing is needed.
\end{itemize}

Note that in this scheme, we always start with the largest possible
candidate value for the time step size.  Subsequently, we may consider
smaller values, but the direction of consideration is always from
larger to smaller, never from smaller to larger.  This guarantees that
we do not run into the flip-flop problem mentioned above.

\section{Numerical Tests}

We present here the results for a gravitational two-body integration.
The relative orbit of the two point masses forms an ellipse with an
eccentricity of $e = 0.99$.  We have chosen a time unit such that the
period of the orbit is $T = 2\pi$.

We have implemented four different integration schemes:

\noindent
0) the original time-symmetric integration scheme described by
\citet{Hut-1995}, where there is a continuous choice of time step size.
This is the approach described in section \ref{subsec:implicit}.
We have used five iterations for each step.

\noindent
1) a block-time-step generalization, with a fixed number of iterations.
This is the approach analyzed in section \ref{subsec:flipflop}.  Here,
too, we chose five iterations for each step.

2) a block time step generalization, with a variable number of iterations.
If after five iterations, the fourth and the fifth iterations still give
a different block time step size, then we choose the smallest of the two.
This recipe avoids flip-flop situations.  It is the approach described
in section \ref{subsec:noflipflop}.

The algorithm described in the next section, \ref{subsec:firsttry}, we
have not implemented here, because it is guaranteed to lead to large
errors in those cases where a new large time step is allowed again
just before pericenter passage.  We therefore switched directly to the
following section:

3) the implementation of our favorite algorithm, where we start with a
truly time symmetric choice of time step, with the restrictions that
we only allow changes of a factor two in the direction of increasing
and decreasing the time step, and that we only allow an increase of
time step on the so-called even time boundaries.  This is the approach
given in section \ref{subsec:secondtry}.

\begin{figure}[ht]
\centering
\includegraphics[width=0.90\textwidth]{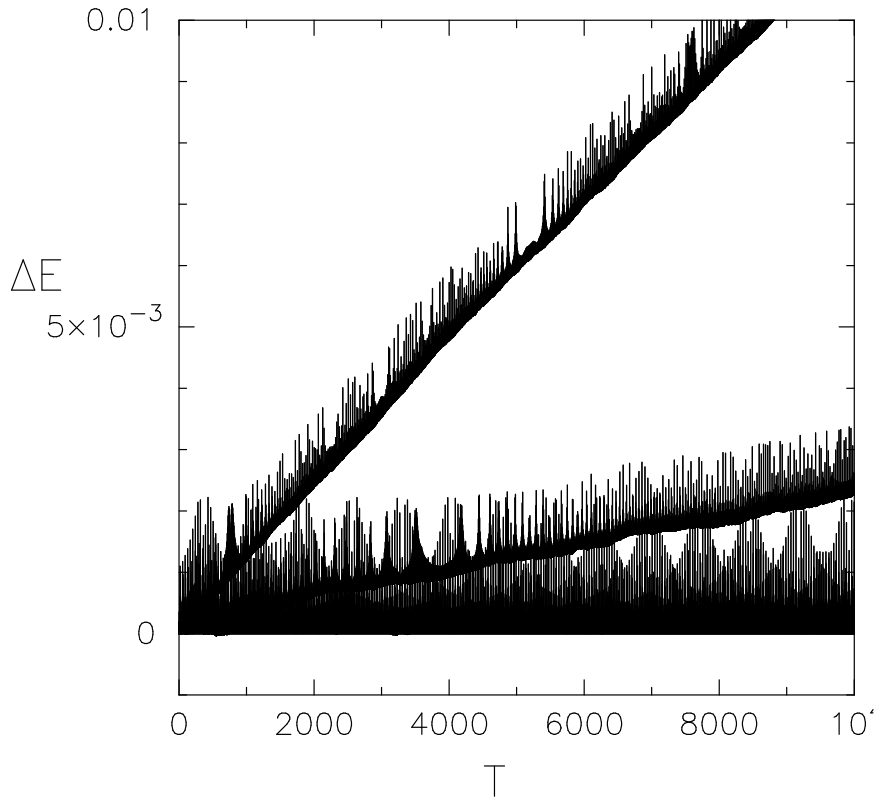}
\caption{Energy errors for a two-body integration of a bound orbit
  with eccentricity $e=0.99$.  The top line with highest slope
  corresponds to algorithm 1, the line with intermediate slope
  corresponds to algorithm 2, and below those the two lines for
  algorithms 0 and 3 are indistinguishable in this figure.}
\label{fig1}
\end{figure}

\begin{figure}[ht]
\centering
\includegraphics[width=0.90\textwidth]{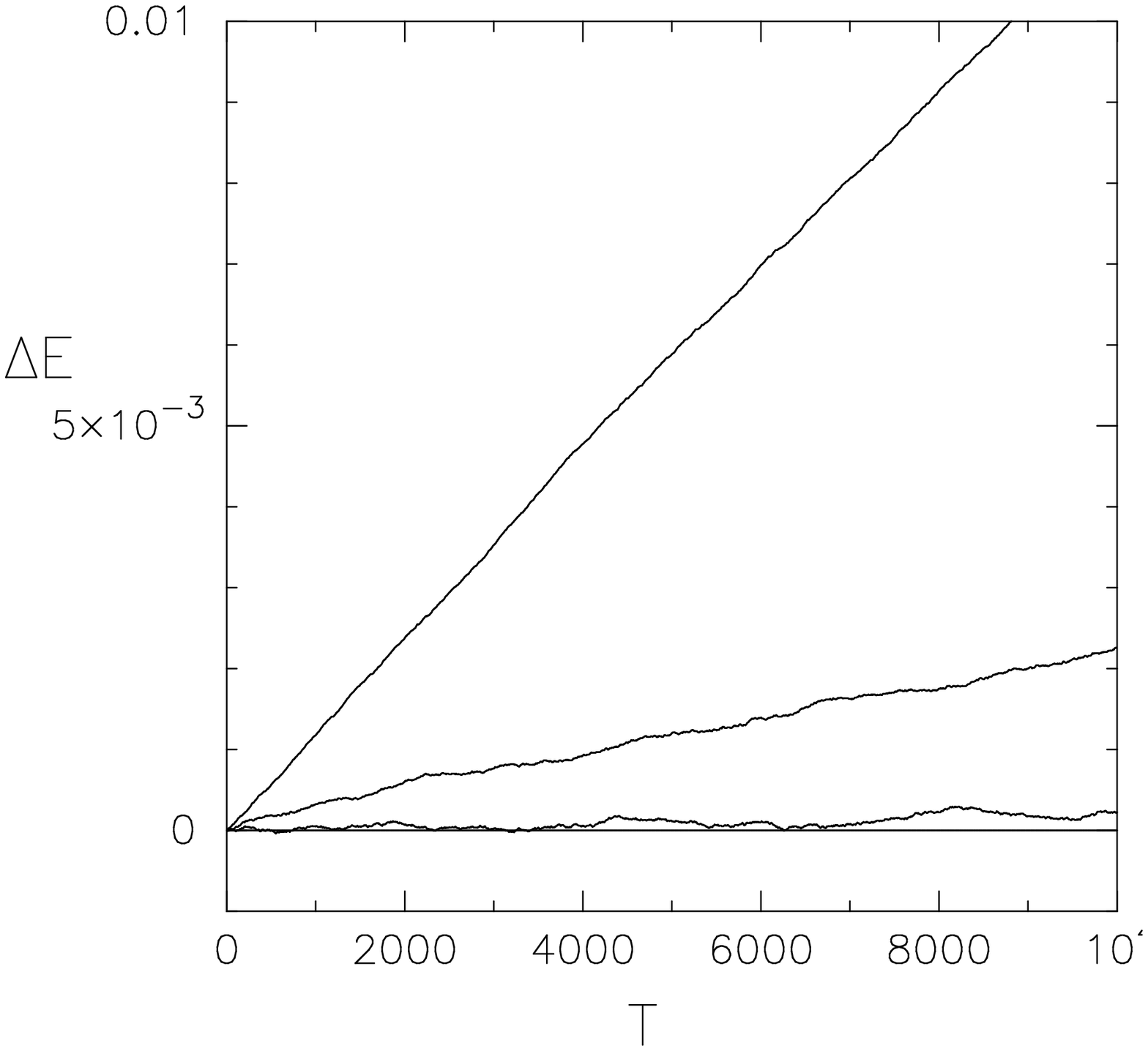}
\caption{The energy errors at apocenter.  The four lines, from top to bottom,
correspond to algorithms 1, 2, 3, and 0.}
\label{fig2}
\end{figure}

In figures \ref{fig1} and \ref{fig2} we show the results of
integrating our highly eccentric binary with these four integration
schemes.  In each case, the largest errors are produced by algorithm
1), smaller errors are produced by algorithm 2), and even smaller
errors appear with algorithm 3).  Finally, algorithm 0) gives the
smallest errors.

Figure \ref{fig1} shows the energy error in the two-body integration
as a function of time.  As is generally the case for time-symmetric
integration, the errors that occur during one orbit are far larger
than the systematic error that is generated during a full orbit.  To
bring this out more clearly, figure \ref{fig2} shows the error only
one time per orbit, at apocenter, the point in the orbit where the two
particles are separated furthest from each other, and the error is the
smallest.

\begin{figure}[ht]
\centering
\includegraphics[width=0.90\textwidth]{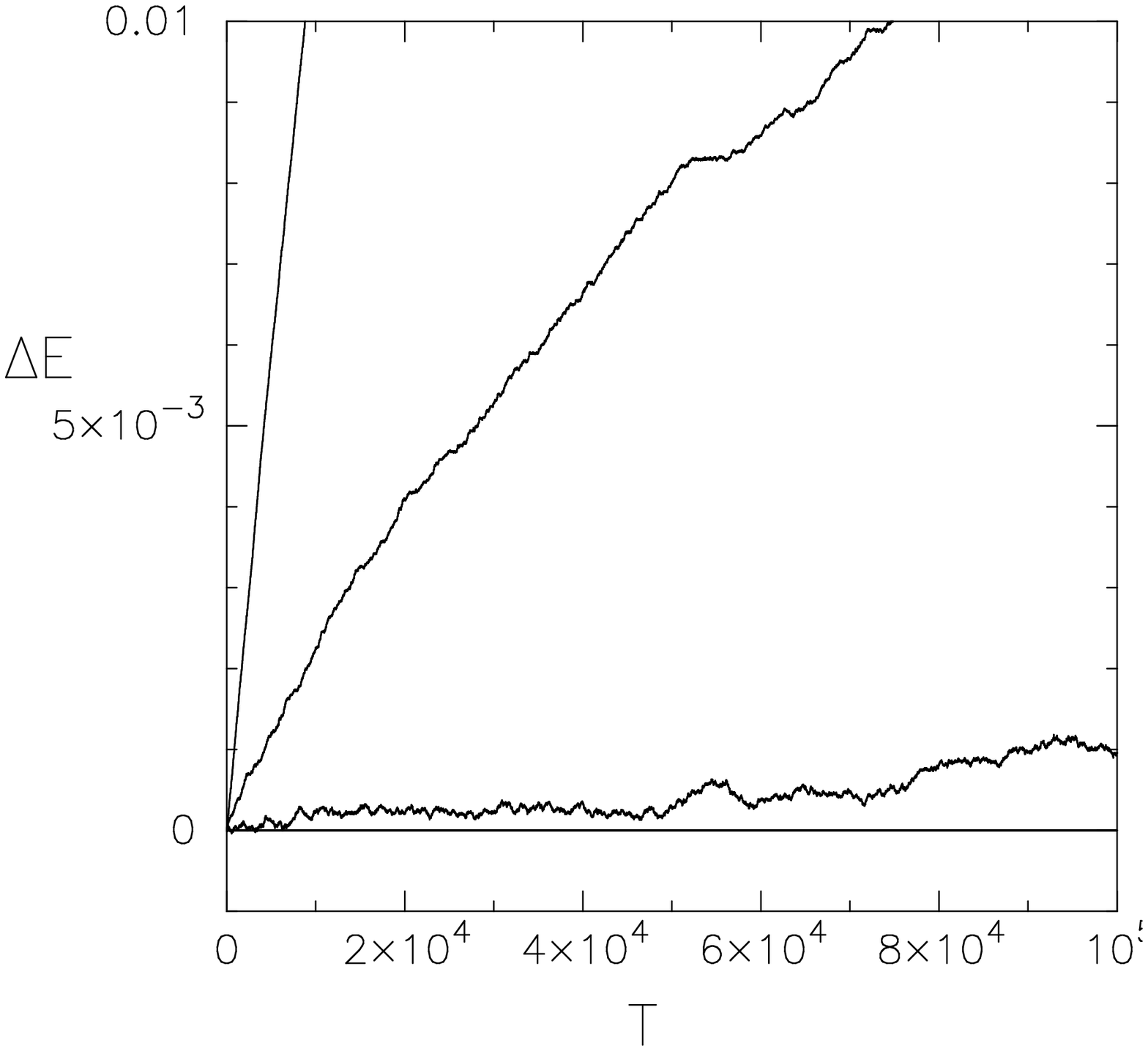}
\caption{Same as Fig. \ref{fig2}, but for a duration that is ten
times longer.}
\label{fig3}
\end{figure}

Finally, figure \ref{fig3} shows the same data as figure \ref{fig2},
but for a period of time that is ten times longer.  In both figures
\ref{fig2} and \ref{fig3}, it is clear that the first two block time
steps algorithms, 1) and 2), both show a linear drift in energy.  This
is a clear sign of the fact that they violate time symmetry.  Note
that in both figures algorithm 3) gives rise to a time dependency that
looks like a random walk.  This may well be the best that can be done
with block time steps, when we require time symmetry.

\section{Summary}

To sum up, we have succeeded in constructing an algorithm for time
symmetrizing block time steps that does not show a linear growth
of energy errors.  As far as we know, this is the first such algorithm
that has been discovered.  We expect this algorithm to have practical
value for a wide range of large-scale parallel N-body simulations.

We acknowledge conversations with Joachim Stadel at the IPAM workshop
on N-Body Problems in Astrophysics, in April 2005, where he presented
a time symmetric version of the preprint by \citet{Quinn-1997} which is
similar to our first attempt at a solution, described in section
\ref{subsec:firsttry}. See also his PhD thesis, ava\-ilab\-le as
 ``www-theorie.physik.unizh.ch/\~\-stadel/downloads/thesis.ps''.
 M.K. and H.S. acknowledge research support from ITU-Sun CEAET grant
5009-2003-03.  P.H. thanks Prof. Ninomiya for his kind hospitality at
the Yukawa Institute at Kyoto University, through the Grants-in-Aid
for Scientific Research on Priority Areas, number 763, ``Dynamics of
Strings and Fields'', from the Ministry of Education, Culture, Sports,
Science and Technology, Japan.

\end{document}